\documentclass[notitlepage]{iopart}
\usepackage{color}
\usepackage{xcolor}
\usepackage[unicode=true, bookmarks=false,breaklinks=false,pdfborder={0 0 1},backref=false,colorlinks=false]{hyperref}
\usepackage{makeidx}
\hypersetup{colorlinks=true,citecolor=blue,linkcolor=blue,filecolor=blue,urlcolor=blue}
\usepackage[english]{babel}
\usepackage[ansinew]{inputenc}
\usepackage{graphicx}
\usepackage{graphics}
\expandafter\let\csname equation*\endcsname\relax	
\expandafter\let\csname endequation*\endcsname\relax 
\usepackage{amsmath}
\usepackage{amsfonts}
\usepackage{amssymb}
\usepackage{epstopdf}
\usepackage{makeidx}
\usepackage{subfigure}
\usepackage{ulem}
\usepackage{pgf}
\usepackage{bm}
\usepackage{tikz} 
\usepackage{titling} 
\usepackage{indentfirst}

\pretitle{\begin{center}\Huge\bfseries}
\posttitle{\par\end{center}\vskip 0.5em}
\preauthor{\begin{center}\Large\ttfamily}
\postauthor{\end{center}}
\predate{\par\large\centering}
\postdate{\par}

\begin{document}
\title{A 50/50 electronic beam splitter in graphene nanoribbons as a building block for electron optics}
\author{Leandro R. F. Lima$^1$, Alexis R. Hern\'andez$^2$, \\ Felipe A. Pinheiro$^2$ and Caio Lewenkopf$^1$}

\address{$^1$ Instituto de F\'{i}sica,
  Universidade Federal Fluminense, 24210-346 Niter\'{o}i, RJ, Brazil}
\address{$^2$ Instituto de F\'{\i}sica,
  Universidade Federal do Rio de Janeiro, \\ Caixa Postal 68528, Rio de
  Janeiro 21941-972, RJ, Brazil}
	
	\ead{leandrolima@if.uff.br, alexis@if.ufrj.br, fpinheiro@if.ufrj.br and caio@if.uff.br}

\date{\today}

\begin{abstract}
Based on the investigation of the multi-terminal conductance of 
a system composed of two graphene nanoribbons, in which one is on top of the other 
and rotated by $60^\circ$, we propose a setup for a 50/50 electronic beam splitter that 
neither requires large magnetic fields nor ultra low temperatures.
Our findings are based on an atomistic tight-binding description of the system and on the 
Green's function method to compute the Landauer conductance.
We demonstrate that this system acts as a perfect 50/50 electronic beam splitter, in which 
its operation can be switched on and off by varying the doping (Fermi energy). 
We show that this device is robust against thermal fluctuations and long range disorder, 
as zigzag valley chiral states of the nanoribbons are protected against backscattering. 
We suggest that the proposed device can be applied as the fundamental element of the 
Hong-Ou-Mandel interferometer, as well as a building block of many devices in electron optics.
\end{abstract}


\submitto{\JPCM}

\ioptwocol

\section{Introduction}
\label{sec:introduction}

The realization of single electron sources \cite{Feve07,Bocquillon13}, where electrons are 
injected on-demand and propagate coherently in a conductor, has open the possibility of 
exploring very interesting new quantum information processing ideas using quantum coherent 
electronics. 
In particular, analogies between single-particle electron and photon propagation triggered 
several beautiful electron transport experiments inspired on optical setups 
\cite{Ji03,Neder06,Roulleau07,Litvin07,Litvin08,McClure09,Zhang09,Henny99,Neder07}.

Currently, the most successful platform for experiments with single-electron sources relies on 
the quantum Hall regime.  
One of the main reasons are the electronic edge states characteristic of the quantum Hall regime.
Those are one-dimensional states with a given chirality, which protects them from electronic 
backscattering \cite{VKlitzing1980,TKNN1982}, improving electronic transport of nanodevices 
and making possible to develop electronic interferometric devices.
Indeed, electronic analogues of many optical devices such as Mach-Zehnder \cite{Ji03,Neder06,Roulleau07,Litvin07,Litvin08},
Fabry-Perot \cite{McClure09,Zhang09}, and Hanbury Brown-Twiss \cite{Henny99,Neder07} 
have been implemented in recent years in Quantum Hall systems. 
These devices are not only important for the investigation of fundamental coherent electronic 
transport but also to applications in quantum computation \cite{Sarkar06}. 
In these devices the mixing of these chiral edge states have been realized using beam splitters 
based on quantum point contracts.
Coherent beam splitters for electrons have been implemented by means of quantum dots with 
normal \cite{Petta10,Oliver1999,Sun2010} and superconducting contacts~\cite{Hofstetter09}. 
However, we emphasize that chiral states in Quantum Hall systems can only occur at high 
magnetic fields and low temperatures, a fact that hampers the development of practical devices 
and technologies.

In order to circumvent these practical limitations, alternative physical concepts have been 
employed in recent years to engineer electronic chiral states. 
Two-dimensional topological insulators that support symmetry protected chiral states 
\cite{Bernevig06,Konig07} have been proposed as suitable candidates, although the robust 
character of their edge states can be experimentally elusive \cite{Knez14,Olshanetsky15}. 

Among the materials exhibiting topological effects that support protected chiral states, graphene 
is one of the most versatile and useful for practical applications \cite{CastroNeto09}. 
Of particular interest are graphene bilayers, which can exhibit interesting topological effects that have been 
proposed to give rise to a so-called valley Hall effect \cite{Xiao2007}, that in monolayer graphene can be achieved by means of lattice deformations \cite{Settnes16,Carrillo16}. Recently, two experimental groups
\cite{Sui2015,Shimazaki2015} succeeded in demonstrating that by changing the Berry curvature 
one can generate and control pure valley currents, the goal of valleytronics, in bilayer graphene.

We adapt  these ideas in a set up of two ``crossed" graphene nanoribbons (GNRs), forming 
a four-terminal device, namely, a graphene bilayer central region coupled to 4 zigzag graphene leads. 
Those are possible to synthesize due to advances in bottom-up nanofabrication techniques, 
such as longitudinal unzipping of carbon nanotubes~\cite{Kosynkin09} and the assembling of 
carbon-based molecules~\cite{Cai10,Jacobberger15}, that have allowed the production of GNRs with high-quality crystallographic edges. 
As a result, the state-of-the art nanofabrication technology can make possible experiments involving 
topological states in GNRs, many of them analogous to the ones conducted in 
quantum Hall systems \cite{Ji03,Neder06,Roulleau07,Litvin07,Litvin08,McClure09,Zhang09,Henny99,Neder07} 
without the application of high magnetic fields.

In this paper we put forward a proposal for a 50/50 electronic beam splitter based on two 
GNRs, one on top of the other forming a four-terminal device as presented 
in what follows, as depicted in Fig.~\ref{fig:system}.
We investigate the electronic transport in this system by means of a microscopic model 
based on the tight-binding approximation and the Landauer conductance formula. 
We demonstrate that it can be applied as a perfect 50/50 electronic beam splitter, in which 
its operation can be switched on and off by varying the electron doping. 
We show that this device is robust against thermal fluctuations and long range disorder, 
since zigzag valley dependent chiral modes of the nanoribbons are protected against backscattering 
\cite{Wakabayashi2007,Wakabayashi2009,Lima12}. 
We suggest that the proposed device can be applied as the fundamental element of the 
electronic analog \cite{Freulon15} of the Hong-Ou-Mandel interferometer \cite{Jachura15}, 
as well as a building block of many devices in electron optics.

This paper is organized as follows. In Sec.~\ref{sec:model} we put forward our model system 
and present the theoretical tools employed in the electronic transport analysis. In Sec.~\ref{sec:results} 
we present the results for the conductance and the Hong-Ou-Mandel interferometer. We present 
our conclusions in Sec.~\ref{sec:conclusions}.

\section{Model and theory}
\label{sec:model}

\begin{figure}[!htbp]
	\begin{center}
		\includegraphics[width=1.0\columnwidth]{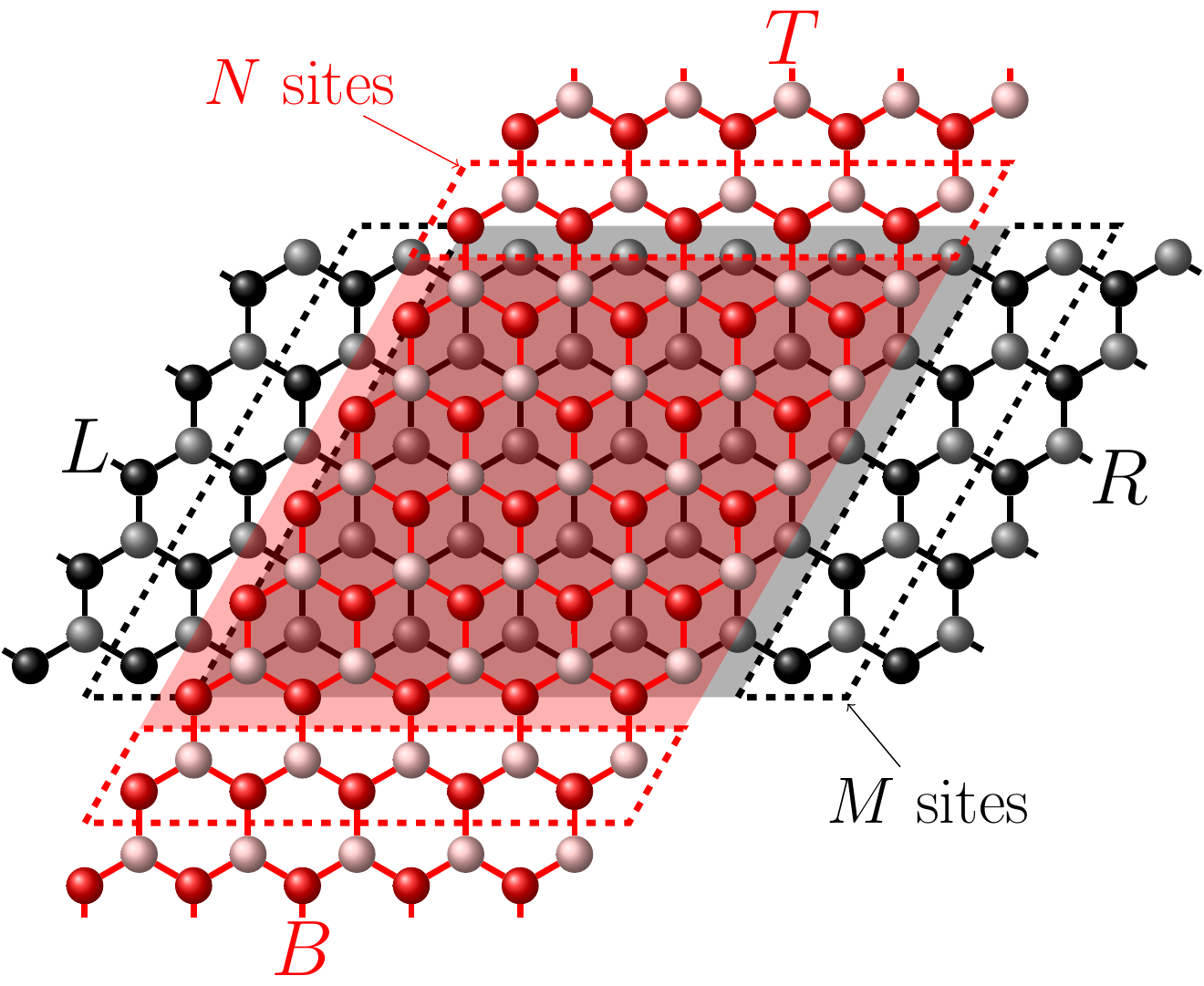}
		\caption{(color online) Sketch of the system. One zigzag GNRs rotated by $60^\circ$ (red) on top of another zigzag GNR on the horizontal direction (black). The number of sites in each ribbon unit cell is $M$ and $N$ for the horizontal and the tilted ribbon, respectively. The shaded area corresponds to the scattering region.
}
		\label{fig:system}
	\end{center}
\end{figure}

The model system we propose is composed of one zigzag GNR along the 
``horizontal'' direction placed underneath another zigzag GNR, tilted by $60^\circ$, as 
depicted in Fig.~\ref{fig:system}.
The relative position and separation between the ribbon planes is such that the atoms 
at their intersection interact as in bilayer graphene with an  {$AB$- or $AA$-like} stacking.

In graphene, $p_z$ orbitals of neighboring carbon atoms hybridize to allow for electronic conduction.
The electronic model Hamiltonian of such a system reads
\begin{align}
H = &-t \sum_{\substack{\left\langle i,j\right\rangle \\ m=1,2}} \left(a_{m,i}^\dagger b^{}_{m,j} + \rm{H.c.}\right) 
\nonumber\\
&+ \sum_{\substack{j \\ m=1,2}} \left(\epsilon_{m,j}^A a_{m,j}^\dagger a_{m,j}^{} + 
\epsilon_{m,j}^B b_{m,j}^\dagger b_{m,j}^{}\right) 
\nonumber\\
&-t_\bot \sum_{\left\langle i,j\right\rangle} \left(a_{1,i}^\dagger b_{2,j}^{} + \rm{H.c.}\right),
			\label{hamiltonian}
\end{align}
where $a_{m,i}^\dagger$ ($b_{m,i}^\dagger$) and $a_{m,i}$ ($b_{m,i}$) are the creation and annihilation operators of the electrons at the $i$-th site of the sublattice $A$($B$). 
The index $m$ labels the top ($m=1$) and the bottom ($m=2$) nanoribbons, whereas $\left\langle ...\right\rangle$ restricts the summations to intra- or interplane first neighbors sites.
The first term accounts for the electronic hopping between nearest neighbors sites in the same plane with an intra-plane hopping integral $t=2.7$ eV. 
The second term includes the onsite energies of the $j$-site in the plane $m$ on the sublattice $A$ ($\epsilon_{m,j}^A$) or $B$ ($\epsilon_{m,j}^B$). 
We study cases where an external electric field is applied in the central region, defined 
as the intersection between ribbons (see Fig.~\ref{fig:system}), so that the potential energy 
in the ribbon planes is constant, $\epsilon_{1,j}^{A,B} = -\epsilon_{2,j}^{A,B} = V/2$. 
 {The interplane potential difference $V$ can be used as a handle} to control the transport properties of the 
central region as it will be shown in the following. Finally, the last term accounts for the interplane 
hopping with hopping amplitude $t_\bot=0.4$ eV \cite{CastroNeto09}.

We use the Landauer approach suited for mesoscopic quantum coherent conductors to calculate 
the transport properties of the four-terminal system depicted in Fig.~\ref{fig:system}.
We assume that the central region, indicated by the shaded area in Fig.~\ref{fig:system}, is 
connected by GNRs to electron reservoirs in equilibrium.
The terminals are denoted by $\alpha=L,R,T,B$, corresponding to left, right, top and bottom 
leads, respectively.

For low bias, the linear conductance from terminal $\beta$ to $\alpha$ is given by the Landauer 
formula \cite{Datta1996}, namely
\begin{align}
G_{\alpha\beta}(\mu) = \frac{2e^2}{\hbar} \int_{-\infty}^{\infty} dE \left(-\frac{\partial f}{\partial E}\right) T_{\alpha\beta}(E)
\end{align}
where $f$ is the Fermi-Dirac distribution and $T_{\alpha\beta}(E)$ is the transmission, 
that is calculated using the Caroli formula \cite{Caroli71,Datta1996}
\begin{align}
	T_{\alpha\beta}(E) = \text{Tr} \left[ \Gamma_\alpha G^r \Gamma_\beta \left(G^r\right)^\dagger\right],
	\label{transmission}
\end{align}
where $G^r(E)$ is the full retarded Green's function of the central region and 
$\Gamma_\alpha$ is the linewidth function of the lead (terminal) $\alpha$.
The notation is standard, see for instance, Ref.~\cite{Lewenkopf13}
for a detailed description of these objects.

Let us briefly describe the numerical implementation of the transmission calculation.
The horizontal and tilted ribbons have $M$ by $N$ sites in the central region.
Hence, owing to the honeycomb lattice structure, the number of sites in each plane is $MN/2$. 
Thus the total number of sites in the central region is $N_{\rm tot} = MN$, which gives the dimension 
of its real space Green's function.
Using the first cell of right and top contacts indicated by the dashed lines in Fig.~\ref{fig:system}, 
we calculate the corresponding surface Green's function by means of a decimation 
technique \cite{Lewenkopf13}.
We obtain the surface Green's functions of the remaining contacts by means of symmetry operations 
on the right and top contact Green's functions.
This simplifies the computation of the retarded self-energies $\Sigma_\alpha^r$ and the line width 
functions $\Gamma_\alpha$. 
We calculate the retarded Green's function of the central region by the (sparse) matrix inversion 
$G^r=(E-H-\Sigma^r)^{-1}$, where the total retarded self-energy is $\Sigma^r=\sum_{\alpha=L,R,T,B}
\Sigma^r_\alpha$.

\section{Results}
\label{sec:results}

Let us now show that by tuning the doping (Fermi energy) and the external electric field on 
the central region, the system introduced in Sec.~\ref{sec:model} can be used to generate a 
50/50 electronic beam splitter. 
Next, we discuss its application as a Hong-Ou-Mandel electron interferometer.

\begin{figure}[!htbp]
\begin{center}
\includegraphics[width=0.99\columnwidth]{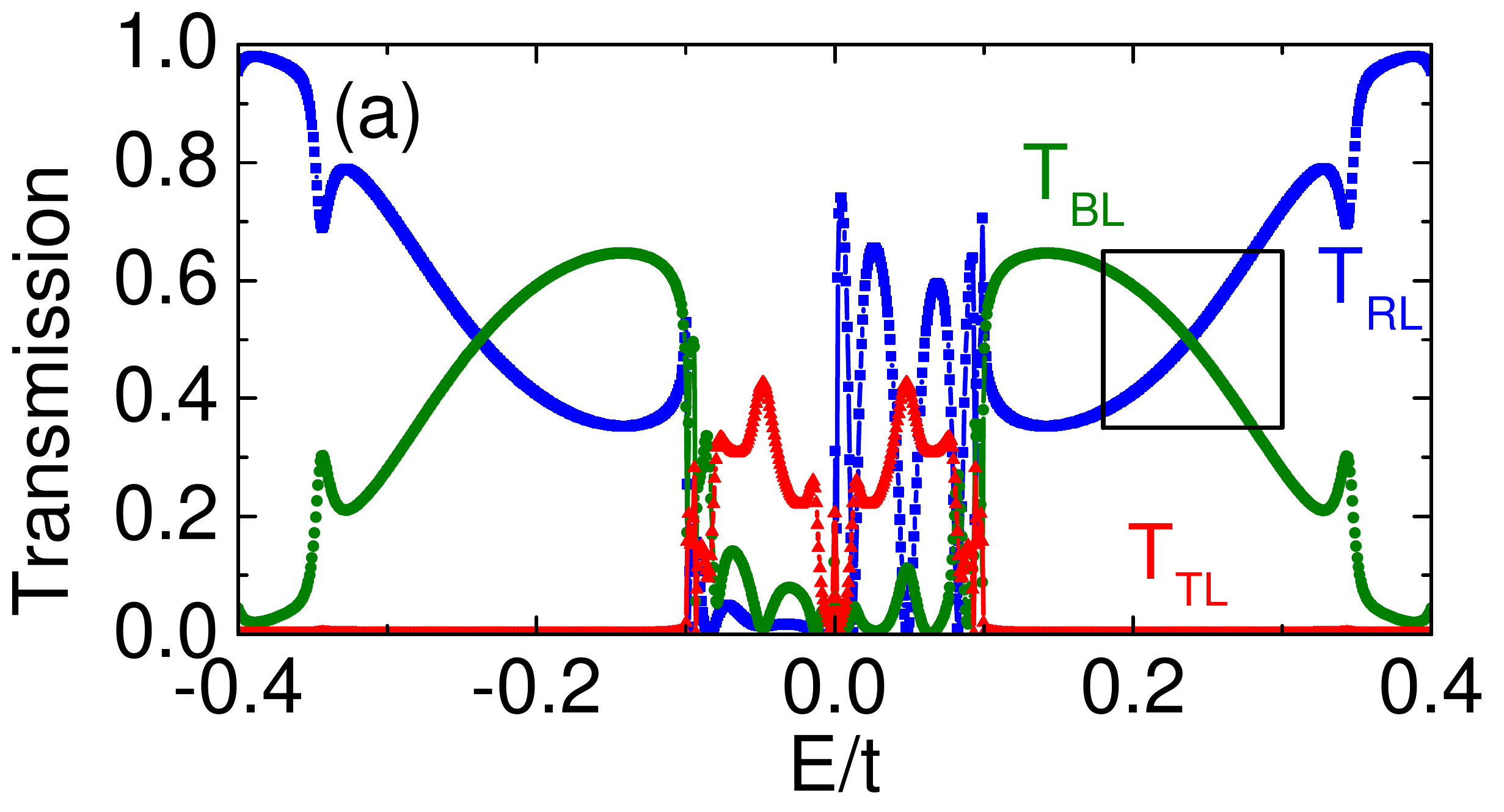}
\includegraphics[width=0.99\columnwidth]{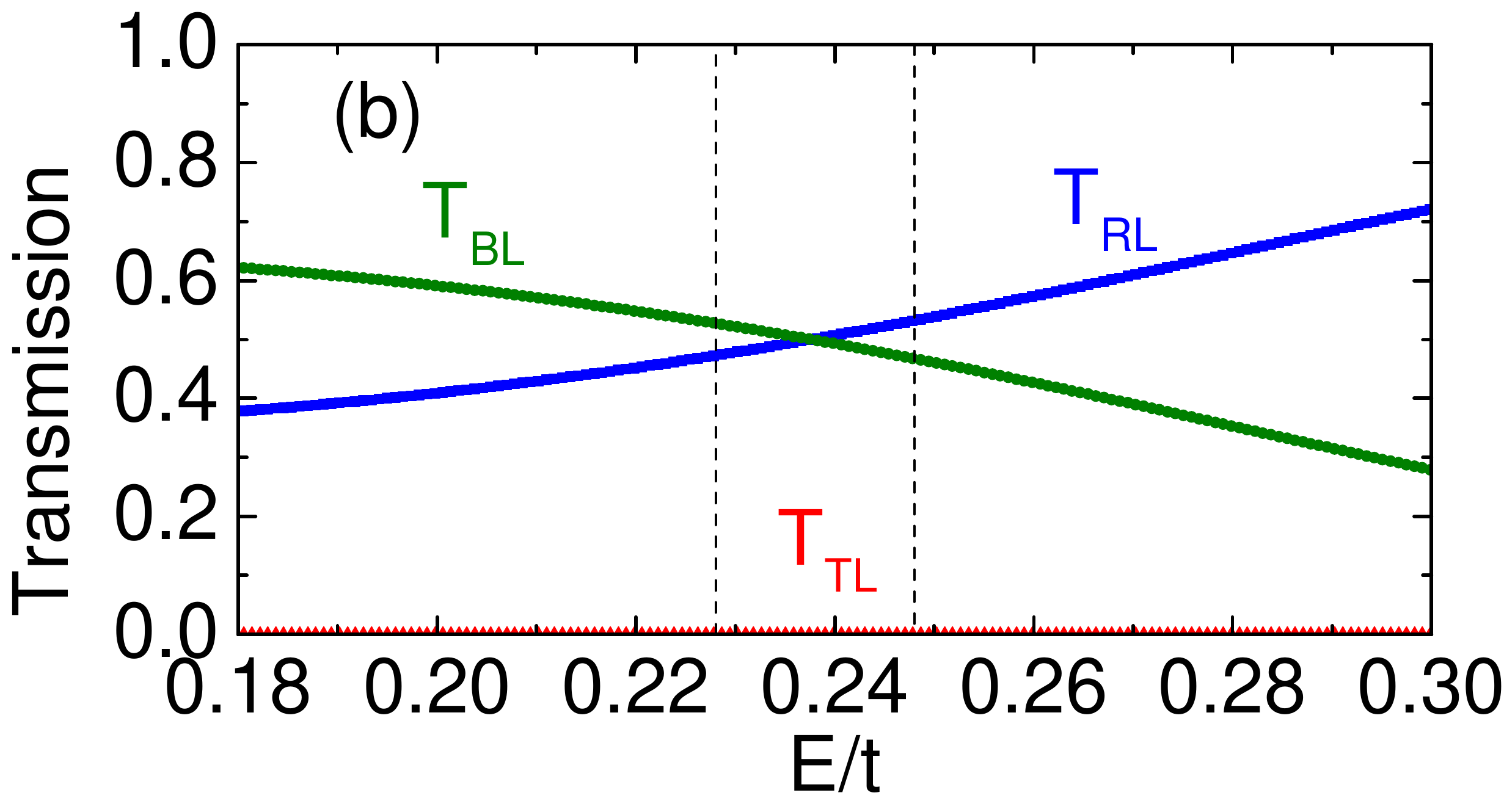}
\caption{{(color online)} (a) Transmissions from left to {top}, right and {bottom} contacts (red, blue and green lines, respectively) as a function of the Fermi energy for $V=\pm 0.1t$. (b) Zoom of the region indicated by the black rectangle in (a). The vertical dashed lines in panel (b) correspond to the energy interval $[-k_BT,k_BT]$ around the crossing point of the transmission curves with temperature $T= 300$ K.
The number of sites in each ribbon cell is $N=M=20$.
}
\label{fig:transmission}
\end{center}
\end{figure}

Figure \ref{fig:transmission} shows the transmission coefficients of electrons injected from the 
left arm of our model system (see Fig.~\ref{fig:system}) for electronic energies close to the charge 
neutrality point.
The panel (a) shows the values of $T_{BL}$, $T_{RL}$ and $T_{TL}$ for $V=\pm 0.1 t$. 
Figure \ref{fig:transmission} reveals that, for a wide energy range, the transmission is almost 
evenly distributed between the right and {bottom} arms. 
This is a result of the valley dependent chiral nature of the modes participating in the electronic 
transport in the zigzag GNRs.
By setting $E \approx 0.24$ eV, transmission in the {top} arm vanishes and the probability of 
transmission in both right and {bottom} arms is $0.5$, as it can be seen in Fig.~\ref{fig:transmission}(b). 
In this regime the system works as a 50/50 electronic beam splitter where the voltage $V$ plays 
the role of a switch on/switch off external parameter. 
Experimentally, the voltage $V$ can be 
controlled by an external electric field and the Fermi energy $E$ {or the doping} is controlled by gate voltages.
{By tuning $E$ and $V$, one can optimize the system operation parameters.} 
Thermal fluctuations, due to temperatures up to $300$ K, do not affect the transport suppression 
to the {top} arm. Indeed, they hardly affect the transmission ratio between right and {bottom} contacts, 
which may reach values lesser than 60/40, as it can be seen in Fig.~\ref{fig:transmission}(b).

In addition, we have verified that if one injects electrons into the {top arm $T$}, the 
transmission follows the same trend of Fig.~\ref{fig:transmission}, where electrons are 
injected in the left arm, provided we set the same values for the potential $V=\pm 0.1t$ 
and the Fermi energy $E \approx 0.24$ eV. In this case, transmission from the {top} 
arm to the left one is {$T_{LT}=0$}; {transmission from the top arm to the right one is 
$T_{RT}=0.5$; and transmission from the top arm to the the bottom one is $T_{BT}=0.5$.}

\begin{figure}[!htbp]
\begin{center}
\includegraphics[width=0.480\columnwidth]{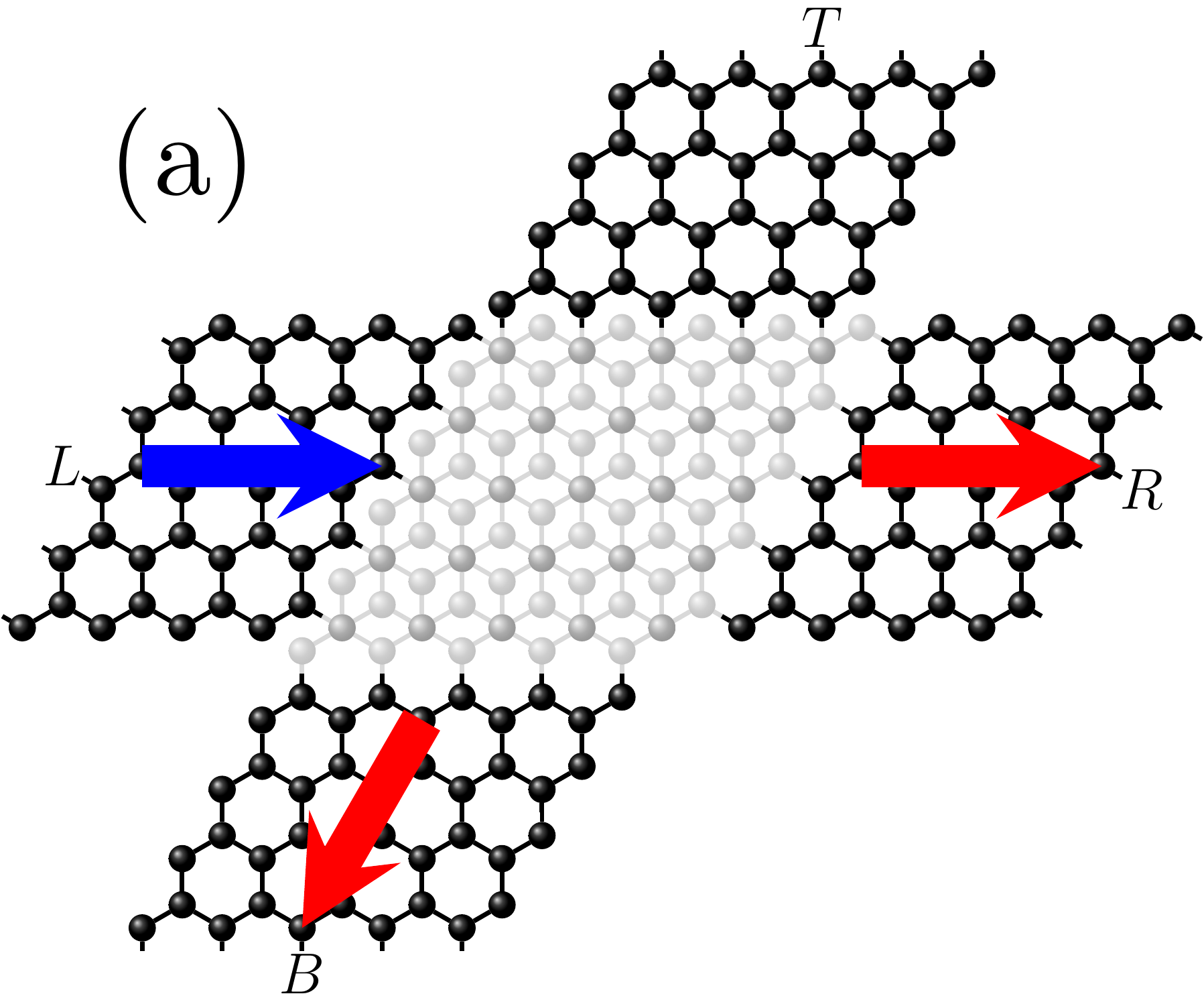}
\includegraphics[width=0.480\columnwidth]{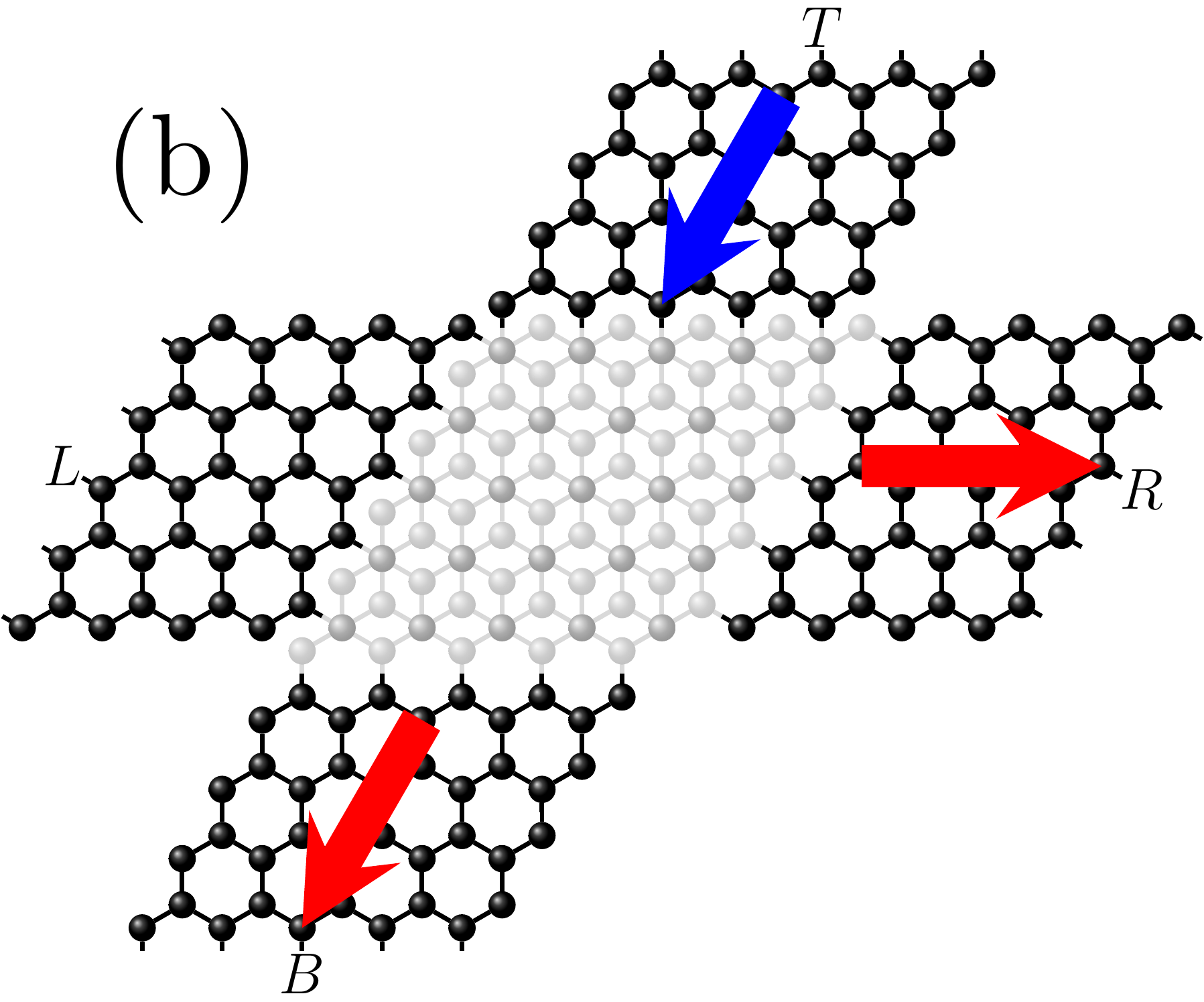}
\includegraphics[width=0.980\columnwidth]{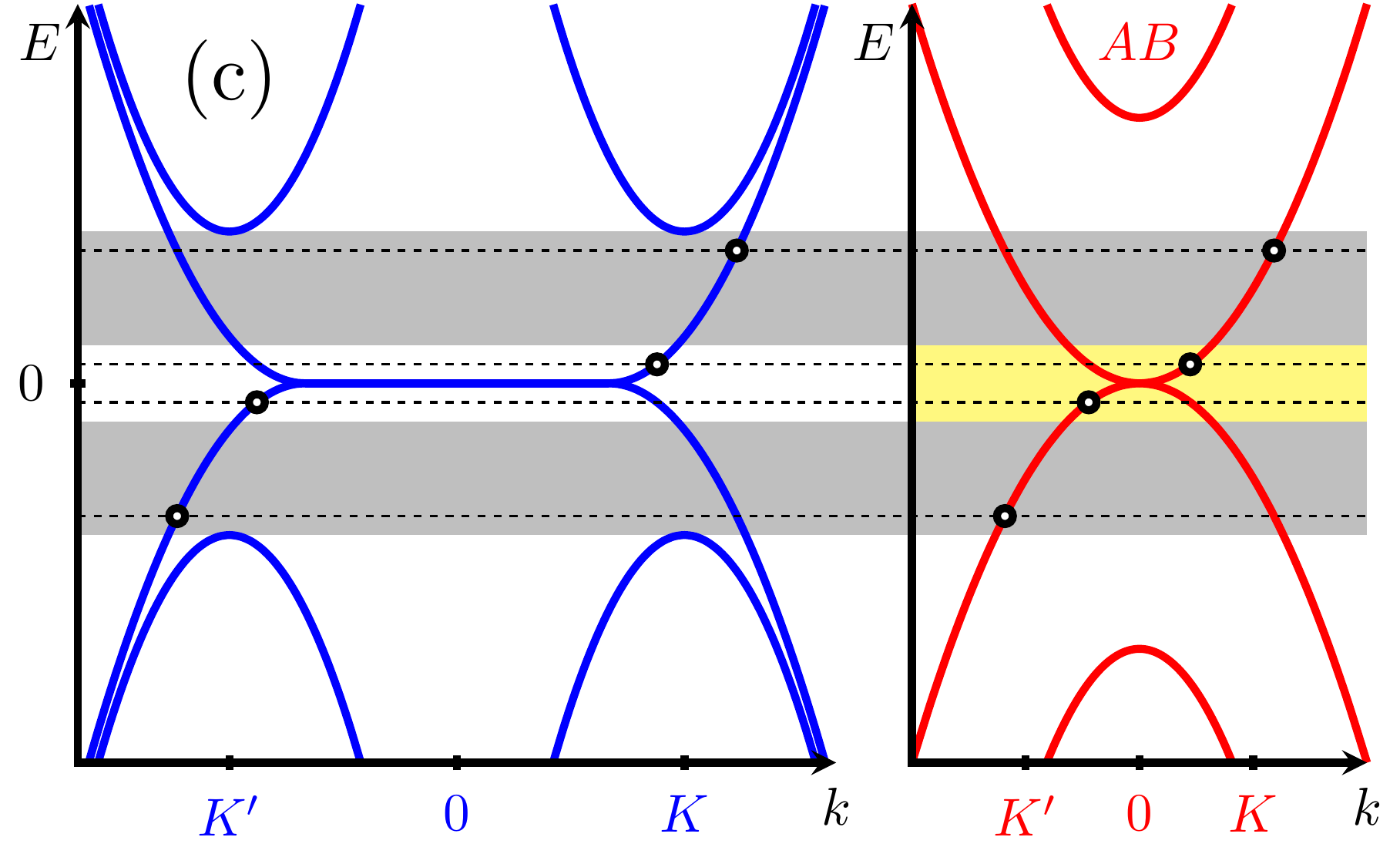}
\caption{{(color online)} Schematic electronic propagation in the 50/50 beam splitter regime and low energy bandstructures of the zigzag GNR and bilayer graphene. Electrons injected from the left $L$ (a) or from the {top arm $T$} (b) can only propagate to {bottom $B$} and right $R$ arms. No propagation between left $L$ and {top $T$} arms is allowed. 
(c) Sketch of the band structures of the injection zigzag GNR and the bilayer central region with $AB$ stacking for $V=0$. 
The shaded areas in gray show the energy windows where valley chirality is preserved. The yellow area corresponds to states with mixed valley chirality. See text.} 
\label{fig:BS_sketch}
\end{center}
\end{figure}

Figure~\ref{fig:BS_sketch} schematically shows the electronic propagation upon injection from 
both left and {top} arms. 
Hence, we have shown that by properly tuning the voltage $V$ and the Fermi energy $E$, 
the system under investigation acts as 50/50 electronic beam splitter, which is a key element 
in many applications in electron optics.

This behavior can be understood as follows: In our proposal the crystallographic zigzag edges 
guarantee that the low energy pair of valley states in each nanoribbon propagate in opposite 
directions \cite{Wakabayashi2007,Wakabayashi2009,Lima12}. The central region composed 
by an AB-stacked bilayer graphene provides interplane scattering that preserves the valley 
index over a wide energy range. 
{In diffusive systems it is standard to use topological arguments:} Since the system geometry does not have inversion symmetry, 
the external perpendicular electric field gives rise to a local (massive) Dirac band structure with 
two valleys characterized by a non-zero Berry curvature with opposite signs (we note that in 
experiments \cite{Sui2015,Shimazaki2015}, inversion symmetry is broken by the substrate 
\cite{Song2015}). Hence, by applying a bias, the electrons at each valley drift in opposite 
directions \cite{Xiao2007}. {Since we address the transport properties using the Landauer formula, there is no explicit in-plane electric field and the reasoning above does not apply \cite{Kirczenow2015}.}

In the system we propose, the electrons are injected in the central region through the zigzag leads. 
As a consequence, for a small doping (or $E_F$) such that only the single open mode operation is 
enabled, the injected electrons through one ribbon, Fig.~\ref{fig:BS_sketch}(a), are always valley polarized \cite{Lima12} and, thus, have a 
preferential deflection direction {when scattered by the bilayer region to the other ribbon.}

By tuning $E$ and the electric field, one can find the optimal operation parameters for 
a 50/50 beam splitter, {see Fig.~\ref{fig:BS_sketch}(a)}. 
In such configuration, the incoming electron from one plane can only 
be scattered to a single state in the other plane with a given propagation direction.
The system shows a chiral symmetry: for electrons injected at the top GNR, one obtains an identical effect as above, illustrated by {Fig.~\ref{fig:BS_sketch}(b)}. 
In our case, for instance, the propagation directions {$T$ to $B$ and $L$ to $R$} have the 
same ``valley chirality", as {verified} by our numerical calculations.

{
Let us discuss the simple case of $V=0$ with the help of Fig.~\ref{fig:BS_sketch}(c).
First we identify that the system in Fig.~\ref{fig:system} is symmetric under rotation around a diagonal axis that takes the arm $L$ ($R$) into the arm $T$ ($B$), interchanging top and bottom nanoribbons.
The low energy electronic transport of each zigzag GNR is characterized by forward and backward moving states having opposite valley chirality \cite{Wakabayashi2007,Wakabayashi2009,Lima12}.
Thus, assuming that the electrons injected in the system from the $L$-arm are $K$-polarized, the forward propagation to $R$ and the scattering to $B$ must be $K$-polarized, while due to symmetry backscattering to $L$ and scattering to $T$ must be $K'$-polarized. 
For $E>0$, away from the charge neutrality point as depicted in the upper gray area of Fig.~\ref{fig:BS_sketch}(c), the electronic states in the bilayer region are fully $K$-polarized and neither backscattering to $L$ nor propagation to $T$ is allowed.
The same arguments apply to the negative energy region $E<0$ corresponding to the lower gray area.
This explains the energy windows with no backscattering and zero transmission between $L$ and $T$ in Fig.~\ref{fig:transmission}. 
On the other hand, for energies $|E|\gtrsim 0$ near the charge neutrality point, the states in the bilayer region lie around $k=0$ having both $K$ and $K'$ polarization components.
Therefore, scattering to any of the four arms is allowed for, enabling backscattering to $L$ and propagation to $T$, as seen in Fig.~\ref{fig:transmission}.
For $V\neq 0$ and/or for $AA$ stacking, the picture is more complicated but the same line of arguments still holds.
}

As an example of application, let us discuss the employment of the electronic beam 
splitter in an alternative implementation of the Hong-Ou-Mandel effect for electrons 
\cite{Neder07,Bocquillon13,Khan14}. 
For bosons, when two indistinguishable particles are incident {on two separate input} 
sides of a 50/50 beam splitter (BS), Bose-Einstein quantum statistics implies that the 
outgoing bosons must leave together in one of the two outputs.
{The coincidence counter placed at the outputs, that detects a signal when two particles 
strike both outputs at the same time, records zero coincidences.}
This effect, first observed for photons \cite{Hong87}, leads to a vanishing coincidence 
for simultaneous {photon injection}, and it is characterized by a dip in the correlation 
function \cite{Jachura15}. 
For electrons, in contrast, Fermi-Dirac statistics implies particle antibunching, {so that 
two identical fermions simultaneously injected at two different terminals are always 
detected in different outputs, leading to a peak in the coincidence count \cite{Khan14}. }

The Hong-Ou-Mandel effect for single fermions was first observed in the one-dimensional 
edge states of quantum Hall systems \cite{Bocquillon13}, although previous measurements 
using continuous electronic beams have been reported \cite{Liu98,Neder07}.
Proposals for the observation of the fermionic Hong-Ou-Mandel effect in graphene also 
exist \cite{Khan14}.

{
We propose that the Hong-Ou-Mandel effect can be verified in our system, depicted in 
Fig.~\ref{fig:system}, by using the terminals $L$ and {$T$} as inputs and the terminals 
$R$ and {$B$} as outputs.
By tuning our system to work as a beam splitter in the regime shown in Fig.~\ref{fig:transmission}(b), 
the electrons injected from terminal $L$ ({$T$}) can be either transmitted to terminal $R$ ({$B$}) with 
transmission $T_{\rm bs}=T_{RL}={T_{BT}}$ or ``reflected'' to terminal {$B$} ($R$) with reflection 
$R_{\rm bs}=1-T_{RL}=1-{T_{BT}}$.
Note that a fine tuning can optimize the system to work as a 50/50 beam splitter where 
$T_{\rm bs}=R_{\rm bs}=0.5$. 
If we introduce a delay time $\delta\tau$ in the injection of one of the electrons and place a 
coincidence counter at the terminals $R$ and {$B$}, the normalized coincidence count $N_c$ 
for our symmetric setup reduces to \cite{Khan14}
}

\begin{align}
	 N_c=T_{\rm bs}^2+R_{\rm bs}^2+2R_{\rm bs}T_{\rm bs}e^{-|\gamma \delta\tau|},
	\label{eqnc} 
\end{align}
where 
$\gamma^{-1}$ is the characteristic time scale of the single electron transistor.
{
In Fig.~\ref{fig:Nc} the coincidence count $N_c$ is shown as a function of the delay 
time $\delta\tau$. It is clear from Fig.~\ref{fig:Nc} that there is a peak at $\delta\tau=0$ 
and for large $\delta\tau$, $N_c$ tends to a flat background, which corresponds to 
uncorrelated transport processes. From Eq.~(\ref{eqnc}) one can verify that the maximum 
difference between the peak and the background occurs for $T_{\rm bs}=R_{\rm bs}=0.5$, 
{\it i.e.} for a 50/50 beam splitter, as it is shown in Fig.~\ref{fig:BS_sketch}.

\begin{figure}[!htbp]
\begin{center}
\includegraphics[width=0.80\columnwidth]{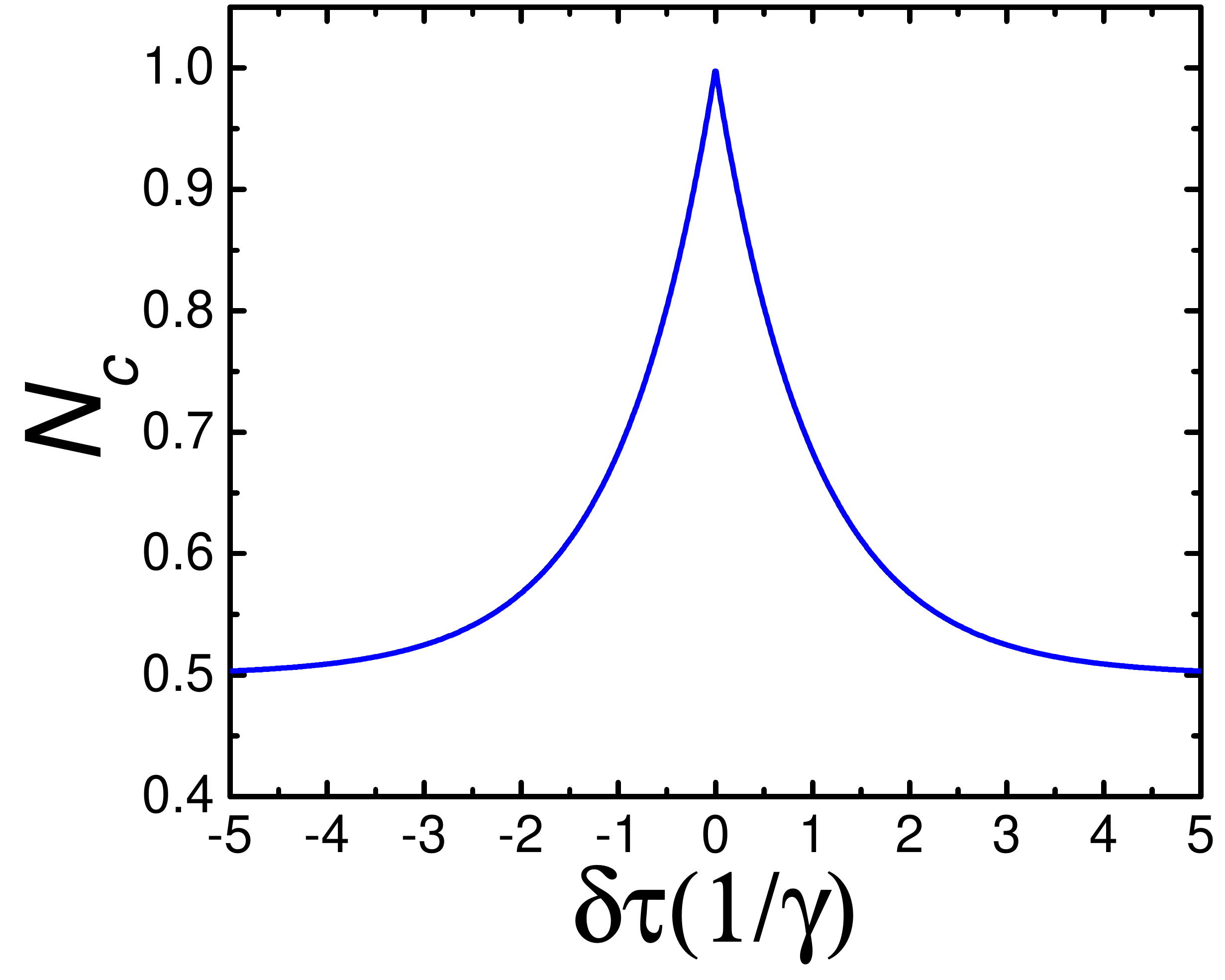}
\caption{ 
(color online) Normalized coincidence count as a function of the time delay between electron sources for a symmetric beam splitter. The time delay $\delta\tau$ is given in units of $1/\gamma$.
}
\label{fig:Nc}
\end{center}
\end{figure}

This enhancement is precisely what can be achieved by the system proposed here with 
realistic parameters. 
In addition, the proposed system, based on the AB stacking of two GNRs, 
preserves the chirality of the states. 
As a result, electrons going from left arm to the right terminal have exactly the same 
chirality of an electron going from the {top to the bottom} arm. 
For this reason the AB staking guarantees that incoming electrons in the left lead will 
only exit the system through the right and {bottom} arms. The same happens for incoming 
electrons in the {top} arm (see Fig.~\ref{fig:BS_sketch}(b)). 

\section{Conclusions}
\label{sec:conclusions}

We have demonstrated that a system composed of zigzag GNRs, 
one rotated by $60^\circ$ on top of another, can function as an ideal 50/50 electronic 
beam splitter. 
We show that the operation of the beam splitter can be switched on and off by varying 
the value of the Fermi energy, which can be achieved by electrostatic gating. 
We also show that the operation of the proposed device is robust against thermal fluctuations.
By making the nanoribbons wider one would change the energy scales of the system trivially. 
We speculate that our main findings still hold.
{Both $AA$ and $AB$ stackings of two GNRs preserve the valley polarization of the states.
We argue that the valley polarization plays a central role in the beam splitter operation.}
{We only presented results for $AB$ stacking, but we have verified that the same conclusions also apply for the $AA$ stacking.}

We emphasize that the transmission for the first modes in zigzag GNRs 
is protected against backscattering due to long range disorder 
\cite{Wakabayashi2007,Wakabayashi2009,Lima12}.
However, short range disorder, either due to edge roughness \cite{Mucciolo09} or resonant 
scattering due to impurity adsorption at the system surface \cite{Duffy2016}, is expected to 
be detrimental for an experimental realization of the proposed beam splitter. We also note that 
the presence of short ranged impurities in the central region give rise to intervalley scattering, 
which destroys the valley polarization.
Recent experiments report an exponential decay of valley currents on the scale of microns 
\cite{Sui2015,Shimazaki2015,Gorbachev2014}.    
In such scenario, finding the conditions of $E$ and $V$ for a proper operation can be challenging.

We suggest that the proposed device can be applied as the fundamental element of the 
Hong-Ou-Mandel interferometer, as well as a building block of many devices in electron 
optics.

\ack

We thank Stephen Power for helpful criticisms.
This work is supported by Brazilian funding agencies CAPES, CNPq and FAPERJ.
F. A. P. acknowledges the financial support of the Royal Society (U.K.) through a 
Newton Advanced Fellowship (ref: NA150208) and the Brazilian agencies CAPES 
(BEX 1497/14-6) and CNPq (303286/2013-0). \\

\bibliographystyle{unsrt}
\bibliography{projectx}

\end{document}